# Circumference, Chromatic Number and Online Coloring

Ajit A. Diwan    Sreyash Kenkre    Sundar Vishwanathan

Department of Computer Science and Engineering,
Indian Institute of Technology Bombay,
Mumbai - 400076, India.

{aad, srek, sundar}@cse.iitb.ac.in

November 7, 2018

**Abstract**

Erdös conjectured that if $G$ is a triangle free graph of chromatic number at least $k \geq 3$, then it contains an odd cycle of length at least $k^{2-o(1)}$ [12, 14]. Nothing better than a linear bound ([3], Problem 5.1.55 in [15]) was so far known. We make progress on this conjecture by showing that $G$ contains an odd cycle of length at least $O(k \log \log k)$. Erdös' conjecture is known to hold for graphs with girth at least 5. We show that if a girth 4 graph is $C_5$ free, then Erdös' conjecture holds. When the number of vertices is not too large we can prove better bounds on $\chi$. We also give bounds on the chromatic number of graphs with at most $r$ cycles of length 1 mod $k$, or at most $s$ cycles of length 2 mod $k$, or no cycles of length 3 mod $k$. Our techniques essentially consist of using a depth first search tree to decompose the graph into ordered paths, which are then fed to an online coloring algorithm. Using this technique we give simple proofs of some old results, and also obtain several simpler results. We also obtain a lower bound on the number of colors an online coloring algorithm needs to use on triangle free graphs.

## 1 Introduction

For a graph $G$, let $\ell(G)$ denote the length of a longest odd cycle (odd circumference), $L(G)$ denote the set of different odd cycle lengths, $\mathcal{C}(G)$ denote the set of different cycle lengths (cycle-spectrum) and let $\chi(G)$ denote the chromatic number (the $G$ will be dropped if the graph is clear). What can we say about $L(G)$, $\mathcal{C}(G)$ or $\ell(G)$ given that $\chi(G)$ is at least $k$? This question is an active area of research. See [2], [3], [5], [12], [14]. The following conjecture is attributed to Erdös.

**Conjecture 1 (Erdös [14],[12])** *Let $G$ be a triangle free graph of chromatic number $k \geq 3$. Then $G$ contains an odd cycle of length at least $k^{2-o(1)}$.*

That $\chi \geq k$ implies the existence of an odd cycle of length at least $k - 1$ is well known ([3] or Problem 5.1.55 in [15]). In fact the bound holds for general graphs (i.e. not necessarily triangle free). A bound of $\Omega(k^2)$ is known for graphs of girth 5. To the best of our knowledge, nothing better is known. Our results give an improvement over these bounds.

The following will be used through out the paper. The circumference of any 2-connected graph is at most $2\ell(G) - 2$ (see Problem 5.1.55 in [15] or [5]). Since the odd cycle lengths possible is a subset of $\{3, 5 \ldots \ell\}$, we have $2|L(G)| + 1 \leq \ell$. The number of different cycle lengths is at most the circumference. By the length of a path we mean the number of edges in that path. The distance between two vertices in a connected graph is the length of a shortest path between them.

## 2 Our Results

We improve upon the linear bound on $\chi$ in terms of $\ell$, for triangle free graphs. In particular, we show that if $G$ is a triangle free graph, then $\chi \leq \frac{15\ell}{\log \log \ell}$. When the number of vertices is not too large, we can improve the above bound. Specifically $\chi \leq O(\sqrt{\frac{\ell}{\log \ell}} \log n)$.

For graphs of girth at least 5, an $\Omega(k^2)$ bound is known (see [2], [12]). Thus the only case where Erdös' conjecture is not known to hold is that of graphs of girth 4. We show that if a girth 4 graph has no $C_5$ present, then Erdös' conjecture holds.

For proving most of the above results, we shall decompose the graph into ordered paths by using a depth first search tree, and then use *online coloring algorithms* on these paths.

We use this technique to prove bounds on the chromatic number when the graph has at most $r$ cycles of length 1 mod $k$ and also when it has at most $s$ cycles of length 2 mod $k$. These results generalize earlier results mentioned in [13].

The use of online coloring leads to questions about how well can we color triangle free graphs online. Halldórsson and Szegedy [4] gave lower bounds on the number of colors used by online coloring algorithms. However their construction may contain triangles. We give a simple construction to show that for every online coloring algorithm and for every $n$, there is a triangle free graph on $n$ vertices, for which the algorithm will need $\Omega(\sqrt{n})$ colors.

The idea of using online coloring after a depth first search ordering has been used before (see [13]). We use this to show simple proofs of linear bounds on $\chi(G)$ in terms of $L(G)$ and $\ell(G)$, and also give very simple proofs of two other results.

# 3 Depth First Search and Online Coloring

In this section we demonstrate the *depth first search*(DFS) (see [1]) and *Online Coloring* (see [6]) method. As is well known, a DFS tree ($T$) is a spanning tree of a connected graph $G$ rooted at a vertex (say $v_0$), with the property that any edge in $G$ either belongs to $T$ or is between an ancestor and a descendant in the tree (also called a *back edge*). For $i \leq j$, let $V[i, j]$ denote the vertices at distance (in $T$) at least $i$, but at most $j$ from the root. In particular, $V[i, i] := V[i]$ denotes the *sphere* or radius $i$ around the root, and $V[i, j] = \cup_{i \leq k \leq j} V[k]$. Since all edges of $G$ connect vertices at different distances from $v_0$ in $T$, each $V[i]$ is an independent set. We begin by giving simple proofs of two well known theorems. The first was proved by Gyárfás [3], answering a question of Erdös and Bollobás. He in fact characterized the graphs for which the inequality below is tight. The second theorem is a result of Erdös and Hajnal (see Problem 5.1.55 of [15]). A strengthening of this was proved in [5].

**Theorem 1 (Gyárfás [3]).**
$$\chi(G) \leq 2|L(G)| + 2 \tag{1}$$

**Proof:** Let $T$ be a DFS tree of $G$, rooted at $v_0$. Let $V_i$ be the vertices of $T$ located at distance $i$ in $T$ from $v_0$. Each $V_i$ is an independent set. An odd(even) level refers to a level $V_i$ with $i$ odd(even). A vertex in level $V_i$ is adjacent to a vertex in level $V_j$, then we say that the levels $V_i$ and $V_j$ are adjacent. No odd(even) level can be adjacent to more than $L(G)$ other odd(even) levels. To see this, note that if $V_i$ and $V_j$ ($i$ and $j$ of same parity) are adjacent, then there are vertices $v_i \in V_i$ and $v_j \in V_j$ that are adjacent in $G$ by a *back edge*, and there is a path of even length $|j - i|$ in $T$ joining $v_i$ to $v_j$. This gives an odd cycle of length $|j - i| + 1$. So if an odd(even) level is adjacent to more than $L(G)$ other odd(even) levels, then $G$ contains more than $L(G)$ distinct odd cycle lengths.

So, in increasing order of levels, greedily assign colors from $\{0, 1 \ldots L(G)\}$ to odd levels. Similarly, in increasing order of levels, greedily assign colors from $\{L(G) + 1, L(G) + 2 \ldots 2L(G) + 1\}$ to even levels. This gives a proper coloring of $G$ using $2L(G) + 2$ colors. In [3], Gyárfás in fact proves a stronger structural result, that the above bound is tight only for graphs containing $K_{2|L(G)|+2}$. ■

The bound of inequality (2) follows easily as follows. Since $3, 5 \ldots, \ell$ are the only possible odd cycle lengths, $L(G) \leq \frac{\ell-1}{2}$. The above theorem thus implies the following result of Erdös and Hajnal. We also give a direct proof, which is a simplification of the above proof.

**Theorem 2 (Erdös-Hajnal (Problem 5.1.55 in [15])).**
$$\chi(G) \leq \ell(G) + 1 \tag{2}$$

**Proof:** As before, let $T$ be a *Depth First Search* (DFS)[1] tree of $G$, rooted at $v_0$. Let $V_i$ be the vertices of $T$ located at distance $i$ from $v_0$ in $T$. By a property of DFS trees, each $V_i$ is an independent set. Assign the color $i \mod (\ell + 1)$ to vertices in $V_i$. We claim that this is a proper coloring of $G$. For suppose that $v$ and $u$ are two vertices adjacent to each other with the same color. Let $v$ and $u$ be at distances $i$ and $j$ from the root $v_0$. Since they are adjacent $i \neq j$. Since they get the same color, $|j - i| = 0 \mod (\ell + 1)$. Hence the unique path in $T$ from $v$ to $u$ along with the $uv$ edge forms an odd cycle of length $\ell + 2$ in $G$, which is a contradiction. In [5] it is proved that the this is tight only for graphs containing a $K_{\ell+1}$. ∎

The coloring procedures we used in the proofs above are examples of *Online Coloring* algorithms, where the algorithm knows the number of vertices a priori. We explain this below (see Kierstead's survey [6] for more details).

An *online graph* $G^\prec$, is a graph $G = (V, E)$ with a total order $\prec$ on its vertices. Suppose $v_1, v_2, \ldots, v_n$ is an ordering of the vertices of $G$ such that $v_i \prec v_{i+1}$ for $i = 1 \ldots n$. An *online coloring algorithm* $\mathcal{A}$ takes as input $G^\prec$, one vertex at a time in the order imposed by $\prec$. It produces a proper coloring of the vertices, but the color assigned to the $i$-th vertex depends only on the graph induced by the vertices $v_1, v_2, \ldots, v_i$. Thus the algorithm irrevocably assigns a color to a vertex based on the vertices it has seen so far. While online coloring algorithms may be randomized, in this paper, *all algorithms considered are deterministic.* Let $\chi_\mathcal{A}(G^\prec)$ denote the number of colors used by $\mathcal{A}$ on $G^\prec$ and let $\chi_\mathcal{A}(G)$ denote the maximum of $\chi_\mathcal{A}(G^\prec)$ across all orderings $\prec$ of its vertices. Let $\Gamma$ be a class of graphs. For every non-negative integer $n$, let $\chi_\Gamma^\mathcal{A}(n)$ denote the maximum of $\chi_\mathcal{A}(G)$ over all $n$ vertex graphs from $\Gamma$ (*most* of the notation follows Kierstead's survey [6]). The following theorem explains how to get bounds for the chromatic number by using DFS trees and online coloring.

**Theorem 3** *Let $\mathcal{A}$ be any online coloring algorithm for graphs in the family $\Gamma$. Any graph $G \in \Gamma$ with a DFS tree $T$ of depth $h$ can be colored with $\chi_\Gamma^\mathcal{A}(h+1)$ colors.*

**Proof:** For convenience, let $f(n)$ denote $\chi_\Gamma^\mathcal{A}(n)$. Let $P_1, P_2, \ldots, P_k$ be the unique paths from the root node $v_0$, to the leaves $w_1, w_2, \ldots, w_k$ of $T$. Color each $P_i$ by presenting it to the procedure $\mathcal{A}$ one vertex at a time, in order of increasing distance from $v_0$. For any $i$ and $j$, let $P_i \cap P_j = \{v_0, v_1, \ldots, v_t\}$ where $v_i$ is at distance $i$ from $v_0$. When $P_i$ or $P_j$ was being colored, the first $t+1$ vertices presented to $\mathcal{A}$ were $v_0, v_1, \ldots, v_t$ in that exact order. Since $\mathcal{A}$ is deterministic, the colors assigned by it to these $t+1$ vertices when coloring $P_i$ is the same as those when coloring $P_j$. Letting each vertex have the color it was assigned by the algorithm, gives a proper coloring of $G$. Since the length of each path was at most $h$, $\mathcal{A}$ needs to use no more than $f(h+1)$ colors. ∎

Thus the use of the DFS tree and online coloring relates the chromatic number to the height of tree.

**Theorem 4** *Let $\mathcal{A}$ be an online coloring algorithm for graphs in the family $\Gamma$. Let $G \in \Gamma$ have odd circumference $\ell$. Then we can color it with $3\chi_{\Gamma,\mathcal{A}}(\ell)$ colors.*

**Proof:** If $G$ has a cut vertex, then we can apply induction. So assume that $G$ is 2-connected. As before, let $f(n)$ denote $\chi_\Gamma^\mathcal{A}(n)$. Using three different palettes of $f(\ell)$ colors, color the subtrees of $T$ induced by $V[i, i+\ell], V[i+\ell+1, i+2\ell], V[i+2\ell+1, i+3\ell]$, by using $\mathcal{A}$ as in the above theorem. Do this for $i = 0, 3\ell+1, 6\ell+1, \ldots$. This yields a proper coloring of $G$. If there are two adjacent vertices $u$ and $v$ that get the same color, then from the above we see that there is a path in $T$ of length at least $3\ell$ between $u$ and $v$. Along with the edge $uv$, this gives a cycle of length $3\ell + 1$. However, any 2-connected graph with odd circumference has a circumference of at most $2\ell - 2$ (see lemma(3) in [5]). This proves the theorem. ■

While online algorithms need not have *a priori* knowledge of the number of vertices or the odd circumference of $G$, since we are primarily concerned with bounds on $\chi$, we assume as and when required that the algorithm is given information about $\ell$ and $n$.

In the next section we use the above theorem to prove sublinear bounds on coloring triangle free graphs. We conclude this section with two simple applications of theorem (3).

**Theorem 5** *If $G$ is triangle free, then the following holds.*

$$\chi \leq \frac{\ell + 3}{2} \text{ if } \ell = 1 \bmod 4 \tag{3}$$

$$\chi \leq \frac{\ell + 5}{2} \text{ if } \ell = -1 \bmod 4 \tag{4}$$

**Proof:** From theorem (3) it suffices to prove that there is a deterministic online coloring algorithm $\mathcal{A}$, which when presented any triangle free graph $P = \{v_1, v_2 \ldots v_n\}$ on $n$ vertices, with each $v_i$ adjacent to $v_{i+1}$ for $i = 1, \ldots, n-1$, in the order $v_1 \prec v_2 \prec \ldots \prec v_n$, uses at most the number of colors mentioned in the statement of the theorem. Algorithm $\mathcal{A}$ works as follows. $\mathcal{A}$ is given the odd circumference $\ell$. Let $\ell'$ be $\ell + 3$ if $\ell = 1 \bmod 4$ and $\ell + 5$ if $\ell = -1 \bmod 4$. Then $\ell'$ is a multiple of 4. We are going to partition the incoming vertices into groups of four, and color each group using two colors. Consider an incoming vertex $v_i$. Let $i' = i \bmod \ell'$. Let $i'$ considered as an integer be at least $4k$ and at most $4k + 3$ for some $k$. If $i'$ is either $4k$ or $4k + 2$, color it with color $k + 1$, else color it with color $k + 2$.

The algorithm produces a proper coloring. Else there is a path $P'$ with vertices $v_x$ and $v_y$ that get the same color. Since we use online graphs, we may assume that $x \prec y$. If $y = x + 1$, then if $x \bmod \ell'$ is $4k$ or $4k + 2$, then $y \bmod \ell'$ is $4k + 1$ or $4k + 3$ respectively. So they get different colors. So suppose that $v_x v_y$ is a back edge. If $y = x + 2$, then $v_y$ and $v_x$ cannot be adjacent as $G$ is triangle free. Then, from the algorithm it follows that there is a subpath of *even* length at least $\ell' - 2$ from $v_x$ to $v_y$. Along with the edge $v_x v_y$, this gives an odd cycle of length at least $\ell' - 1$, i.e. at least $\ell + 2$. This is a contradiction.

For counting the number of colors, note that the vertices are partitioned into $\frac{\ell'}{4}$ groups of size four, and two colors are used on each group. Thus the number of colors used is $2\frac{\ell'}{4} = \frac{\ell'}{2}$, which is the statement of the theorem. ∎

In the following theorem, the graph need not be triangle free.

**Theorem 6** *If $p$ is the length of a longest path in a graph $G$ with clique number $\omega$, then*

$$\chi \leq \frac{p+\omega}{2} \qquad (5)$$

**Proof:** From theorem (3) it suffices to show that there exists an online coloring procedure that uses at most $\frac{n+\omega}{2}$ colors when presented a graph on $n$ vertices and clique number at most $\omega$. We show that the *First Fit* (FF) procedure suffices for this. In the FF algorithm the colors correspond to integers, and each incoming vertex is assigned the smallest color to which it is not adjacent. Suppose a graph $P$ on $n$ vertices and clique number at most $\omega$ is presented online to the FF algorithm, and let $x$ be the number of colors used by FF to color these $n$ vertices. Let $t$ be the number of colors with only one vertex in their color classes, and let $u_1, u_2 \ldots u_t$ be the corresponding vertices. Wlog suppose that $u_i$ occurs before $u_{i+1}$ in the online order, for $i = 1 \ldots t-1$. Since exactly $x$ colors are used, we have

$$t + 2(x-t) \leq n \qquad (6)$$

This implies that $x \leq \frac{n+t}{2}$. Since we are using First Fit, and each $u_i$ is the only vertex of its color, every $u_i$ is adjacent to all $u_j$ for $j$ less than $i$. Thus $P[u_1, u_2 \ldots u_t]$ is the clique $K_t$. Thus $t \leq \omega$, giving $x \leq \frac{n+\omega}{2}$, which proves the theorem. ∎

## 4 Sublinear Bounds on $\chi$

In this section we prove the sublinear bounds on $\chi$ for triangle free graphs, as a function of $\ell$.

**Theorem 7** *If $G$ is a triangle free graph, then $\chi \leq \frac{15\ell}{\log \log \ell}$.*

**Proof:** By theorems (3) and (4) it suffices to show that there is an online algorithm $\mathcal{A}$ with $\chi_\Gamma^\mathcal{A}(n)$ at most $O(\frac{n}{\log \log n})$, where $\Gamma$ is the class of all triangle free graphs. We use the following result.

**Theorem 8 (Lovász, Saks and Trotter [7])** *Every online graph $G$ can be partitioned deterministically into sets $D_1, D_2, \ldots, D_d$ and $C_1, C_2, \ldots, C_r$, where each $D_i$ is independent and each $C_i$ is contained in the neighborhood of some vertex; if we take $d = n/\log \log n$, then $r = 4n/\log \log n$.*

Thus, if $G$ is a triangle free graph, then each $C_i$ in the above theorem is independent as it lies in the neighborhood of a vertex. Thus $\mathcal{A}$ uses the above online partitioning algorithm to partition the online graph into at most $5n/\log\log n$ independent sets, where each independent set corresponds to a color class. Thus $\chi_\Gamma^{\mathcal{A}}(n) \leq \frac{5n}{\log\log n}$. Now, using Theorem 4, we can color any triangle free graph with $\frac{15\ell}{\log\log \ell}$ colors. ∎

Thus, if a triangle free graph has chromatic number at least $k \geq 3$, then it contains an odd cycle of length at least $\Omega(k \log \log k)$. This is far from the $\Omega(k^2)$ bound in Erdös' conjecture, but is certainly an improvement to $\Omega(k)$ bounds known so far. A quadratic bound is known for graphs with girth at least 5. We give a simple proof below.

**Theorem 9** *If $G$ is a graph of girth at least 5, then $\chi \leq O(\sqrt{\ell})$.*

**Proof:** A result weaker than the above two can be proved using online coloring. We show that there is an online coloring algorithm $\mathcal{A}$ with $\chi_\Gamma^{\mathcal{A}}(n) \leq 2\sqrt{n}$, where $\Gamma$ is the class of graphs of girth at least 5. Let $k = 2\sqrt{n}$ and let $\mathcal{A}$ be the first fit algorithm. Suppose the $n$-th vertex is adjacent to vertices of each of the $k$ colors, and let $u_1, u_2 \ldots u_k$ be these vertices, where $u_i$ has color $i$. Then each vertex $u_i$ is adjacent to vertices $w_1^i \ldots w_{i-1}^i$ of each of the colors $1, 2 \ldots i-1$. Since the girth is at least 5, no $w_i^j$ equals another $w_i^{j'}$ for $j \neq j'$. Hence there are at least $1 + 2 + 3 + \cdots + k$ vertices, which exceeds $n$. Since this is impossible, the theorem follows. ∎

We note that since the circumference can be at most $2\ell - 2$, the number of different cycle lengths is at most $2\ell - 2$. Thus, theorem (2.5) of [12] and theorem (1) of [2] are both strengthening of the above theorem.

For triangle free graphs that contain a $C_4$, no such quadratic bound is known. However, for triangle free graphs that may contain a $C_4$ but no $C_5$, we have the following bound.

**Theorem 10** *If $G$ be a triangle free graph with no $C_5$ subgraph, then $\chi \leq 6\sqrt{\ell}$.*

**Proof:** Let $\Gamma$ be the class of triangle free graphs with no $C_5$. By theorems (4) and (3) it suffices to find an online coloring algorithm $\mathcal{A}$ such that $\chi_\Gamma^{\mathcal{A}}(n) \leq 2\sqrt{n}$. We can use the following result of Kierstead.

**Theorem 11 (Kierstead [6])** *For every positive integer $n$, there exists an on-line coloring algorithm $B$ such that $\chi_B(G) \leq 2\sqrt{n}$, for any graph on $n$ vertices that contains neither $C_5$ nor $K_3$.*

If we let $\mathcal{A}$ be the algorithm of the above theorem, then $\chi_\Gamma^{\mathcal{A}}(n) \leq 2\sqrt{n}$. Hence graphs that are $K_3$ and $C_5$ free are $6\sqrt{\ell}$ colorable. ∎

It is known that if $G$ is a triangle free graph on $n$ vertices, then $\chi(G) \leq O(\sqrt{\frac{n}{\log n}})$ (see [10]). When the circumference is small, we can get a better bound on $\chi(G)$.

**Theorem 12** *If $G$ is a triangle free graph on $n$ vertices, then $\chi \leq O(\sqrt{\frac{\ell}{\log \ell}} \log n)$.*

**Proof:** As before, do a DFS on $G$. As in theorem (4) split the DFS tree into levels of height $\ell$. It suffices to color a height $\ell$ tree with $O(\sqrt{\frac{\ell}{\log \ell}} \log n)$ colors. So let $T$ be a DFS tree of height $\ell$, and suppose it has $k$ leaves. Trace the following path $P$ in the tree starting from the root. At each vertex $v_i$ we encounter, follow that path to the subtree of $v_i$ with the largest number of leaves. Clearly, each subtree of $T - P$ has at most half the number of leaves as those in $T$. We will color $P$ with $\sqrt{\frac{\ell}{\log \ell}}$ colors. We then recurse this procedure on each of the subtrees of $T - P$, where the same palette of colors is used for two subtrees with roots at the same depth from the root in $T$ (since $T$ was a DFS tree, there are no edges between such subtrees). Since the number of leaves in each subtree is at least halved at each step, there can be at most $\log n$ levels in the recursion.

Since $T$ has height at most $\ell$, each of the paths generated in the recursion has at most $\ell$ vertices. Since $G$ is triangle free, each such path can be colored with $O(\sqrt{\frac{\ell}{\log \ell}})$ colors. Since there are at most $\log n$ levels of recursion, $G$ can be colored with at most $O(\sqrt{\frac{\ell}{\log \ell}} \log n)$ colors (using theorem (4)). ■

Note that if every path in $G$ is $k$ colorable, then each of the paths we generated the above proof can be colored with $k$ colors. Thus we have the following corollary (also see [11] for a similar result).

**Corollary 1** *If every path in a graph $G$ is $k$ colorable, then $G$ is $3k \log n$ colorable.*

## 5 Coloring With Excluded Cycle Lengths

In this section we prove some results that follow directly from the above technique of using DFS and online coloring. Tuza and Toft [13] observed that every graph with chromatic number greater than $k$, has a cycle of length divisible by $k$. This is equivalent to saying that if a graph has no cycle of length 0 mod $k$, then the graph is $k$ colorable. Tuza also proves that if $G$ has no cycle of length 1 mod $k$ then $G$ is $k$ colorable. We extend this to the case when there are at most $r$ distinct cycles of lengths one more than a multiple of $k$. When $k = 2$, we recover the result of theorem (1).

We will use the following notation: if $u$ and $v$ are two vertices in a given tree $T$, then the unique path in $T$ between them will be denoted by $P_{uv}$.

**Theorem 13** *Let $k$ be a non negative integer. If a graph $G$ has at most $r$ cycles of different length 1 mod $k$, then $\chi(G) \leq rk + k$.*

**Proof:** Let $T$ be a DFS tree of $G$. Let $\mathcal{A}$ be the first fit online coloring algorithm. For $i = 0 \ldots k - 1$, let $W_i$ be the vertices at distance $i$ mod $k$ from the root vertex $v_0$. We feed each $W_i$ to $\mathcal{A}$ as follows: the *next* vertex is that among the uncolored vertices of $W_i$, closest in $T$ to $v_0$ (ties are broken arbitrarily).

We claim that each $W_i$ can be colored with at most $r+1$ colors. For suppose $\mathcal{A}$ is coloring the vertex $v$ in $W_i$. Let $w_1, w_2, \ldots, w_t$ be the colored vertices of $W_i$ adjacent to $v$, through the edges $e_1, e_2, \ldots, e_t$. Since $v$ and each $w_j$ is in $W_i$, each $P_{vw_j}$ has length $0 \bmod k$. Thus, each of the cycles $P_{vw_j} e_j$ for $j = 1, \ldots, t$ have length $1 \bmod k$. Thus $k \leq r$, and so $\mathcal{A}$ uses no more than $r+1$ colors on each $W_i$.

By using a different palette for coloring each $W_i$, we get a $(r+1)k$ coloring of $G$. ∎

We extend this for the case when the graph has at most $s$ cycles of length $2 \bmod k$.

**Theorem 14** *Let $k$ be a non negative integer. If a graph $G$ has at most $s$ cycles of different lengths $2 \bmod k$, then $\chi(G) \leq sk + k + 1$.*

**Proof:** Let $T$ be a DFS tree of $G$ and let $\mathcal{A}$ be the first fit algorithm. As in theorem (3) let $P_1, P_2, \ldots, P_k$ be the unique paths from the root node $v_0$, to the leaves of $T$. Color each $P_i$ by presenting it to $\mathcal{A}$ one vertex at a time, in order of increasing distance from $v_0$.

Consider a vertex $v$ at a distance $i \bmod k$ from $v_0$. Let $u_1, u_2, \ldots, u_t$ be its ancestors at distance $(i-1) \bmod k$ from $v_0$. Let $e_1, e_2, \ldots, e_t$ be the edges through which $v$ is adjacent to $u_1, \ldots, u_k$ respectively. We may assume that $u_1$ is the immediate ancestor of $v$ in $T$. Then $e_1$ is a tree edge and $e_2, \ldots, e_t$ are back edges. Clearly, all the $u_i$'s lie on the path $P_{v_0 v}$ in $T$ from $v_0$ to $v$. Since the $u_i$'s are at the same distance modulo $k$ from $v_0$, the paths in $T$ between them have lengths $0 \bmod k$. Then for each $u_i$ with $i = 2, \ldots, t$, each of the $t-1$ cycles, $e_i, P_{u_i u_1}, e_1$, have length $2 \bmod k$. So $t \leq s+1$. Hence if $v$ is adjacent to at least $sk + k + 1$ of its ancestors, then at least $sk + k + 1 - (s+1) = (s+1)(k-1) + 1$ of these are at distances different from $(i-1) \bmod k$. Then there are at least $s+2$ of these adjacent ancestors (say $w_1, \ldots, w_{s+2}$), which are at the same distance modulo $k$ from $v_0$. For $j = 1, \ldots, s+2$, let $f_j$ denote the edge $vw_j$. Then $f_1, P_{w_1 w_j}, f_j$ for $j = 2, \ldots, s+2$ are $s+1$ cycles of length $2 \bmod k$. Thus $v$ can be adjacent to at most $sk + k$ of its ancestors. So $\mathcal{A}$ uses at most $sk + k + 1$ colors on $G$. ∎

We have the following obvious corollary.

**Corollary 2** *Let $k$ be a non negative integer. If a graph $G$ has no cycles of length $2 \bmod k$, then $\chi(G) \leq k + 1$.*

When $k = 2$, then the condition of the theorem is equivalent to stating that the graph has at most $s$ cycles of different *even* lengths. We thus recover the following bound first proved by Mihók and Schiermeyer [9].

**Corollary 3 (Mihók and Schiermeyer [9])** *If a graph $G$ has at most $s$ even length cycles of different lengths, then $\chi(G) \leq 2s + 3$.*

Mihók and Schiermeyer [9] also gave a structural result to characterize the graphs for which the above inequality if tight.

If $G$ has no cycles modulo 3, we get a $2k$ bound. We suspect that this is far from tight.

**Theorem 15** *If $G$ is a graph with no cycles of length $3 \mod k$, then $G$ is $2k$ colorable.*

**Proof:** Let $T$ be a DFS tree of $G$ and let $V[i]$ be the vertices at distance $i$ from the root $v_0$. Let $W_i$ be the vertices at distance $i$ modulo $k$ from $v_0$ i.e.

$$W_i = \cup_{j=i \bmod k} V[j] \tag{7}$$

Let $H_i$ be the subgraph of $G$ induced by $W_i$. We claim that each of the $H_i$'s are acyclic. Suppose that $C$ is a cycle in $H_i$. Let $u$ be a vertex of $C$, furthest from the root, and let $v$ and $w$ be the two neighbors of $u$ in $C$. We assume without loss of generality that $v$ is an ancestor of $w$. Let $e_1$ and $e_2$ denote the edges $uv$ and $uw$ respectively. Let $x$ be the neighbor of $w$, other than $u$ in $C$. Suppose $x$ is an ancestor of $w$. Since $v$ and $x$ lie in $H_i$, $P_{vx}$ has length $0 \mod k$. Then the cycle $e_1$, $P_{vx}$, the edge from $x$ to $w$, and the edge $e_2$ give a $3 \mod k$ cycle in $G$. So suppose that $x$ is a descendant of $w$. Since $C$ is a cycle and no cross edges are allowed in $T$, there is an edge $e_3$ of $C$, joining a descendant of $w$ (say $z$), with an ancestor of $w$ (say $y$). Since $v, w, y, z$ are in $H_i$, the paths $P_{vz}$ and $P_{yw}$ have lengths $0 \mod k$. Then the cycle consisting of the edges $e_1$, the path $P_{vz}$, the edge $e_3$, the path $P_{yw}$, and the edge $e_2$, give a cycle of length $3 \mod k$ in $G$. Thus each $H_i$ is acyclic. Now we can color each $H_i$ with a different set of 2 colors to get a $2k$ coloring of $G$. ∎

A special case of the above is when $k$ equals 3. Then there are no cycles of length $0 \mod k$. We thus have the following result.

**Corollary 4** *If $G$ is a graph with no cycle of length $0 \mod 3$, then $G$ is $6$ colorable.*

## 6  Lower Bound For Online Coloring

Let $\Gamma$ be the class of triangle free graphs. We have seen that if there is an online coloring algorithm $\mathcal{A}$ with $\chi_\Gamma^\mathcal{A}(n) \leq f(n)$, then for each $G \in \Gamma$ $\chi(G) \leq 3f(\ell)$. This raises the question of how good can $\chi_\Gamma^\mathcal{A}(n)$ be? We show that it cannot be less than $\sqrt{n}$.

**Theorem 16** *Let $\mathcal{A}$ be any online coloring algorithm. Then for each $n$, there is an $n$ vertex graph $G$ in $\Gamma$ with an ordering $\prec$, such that $\mathcal{A}$ takes at least $\sqrt{n}$ colors to color the online graph $G^\prec$.*

**Proof:** Let $S_1 \ldots S_k$ be the $k$ subsets of $[k] = \{1, 2 \ldots k\}$, where $S_i = [k] - \{i\}$. We assign colors from $[k]$ to the vertices, while the algorithm $\mathcal{A}$ will assign the vertices to *bins*. Initially, set the *current color set* to be $S_1$. When the next vertex $v$ arrives, make it adjacent to all vertices colored $i$, where $S_i$ is the current color set. Let the algorithm assign $v$ to the bin $B$. Choose a color in $S_i$ that is not in $B$, and assign it to $v$. If $B$ gets all the $k$ colors, change the current color set to $S_{i+1}$. Run this procedure for $k^2$ vertices.

Since the neighborhood of each vertex is independent, the graph is triangle free. It is clearly $k$ colorable. Also, since each bin can contain no more than $k$ vertices, there are at least $k$ bins. Thus any online procedure will use $k$ i.e. $\sqrt{n}$ colors. ∎

This example is tight for the *First Fit* coloring algorithm. The graph generated for the First Fit algorithm has maximum degree of $k$, and hence is $O(k/\log k)$ colorable by Johansson's result ([8]). Note that we reveal the color we assign to a vertex to the algorithm. This idea is also used in the construction of Halldórsson and Szegedy [4], which holds for general graphs (i.e. may contain triangles).

# 7 Conclusion

There are triangle free graphs with chromatic number $\Omega(\sqrt{\frac{\ell}{\log \ell}})$. While Erdös' conjecture states that there cannot be a triangle free graph with chromatic number greater than $\sqrt{\ell}$. We have proved an upper bound of $\frac{15\ell}{\log \log \ell}$. The exact values remain unresolved.

Another question is how well can an online algorithm perform on a triangle free graph? We have shown that it can do no better than $\sqrt{n}$. If there is an online coloring algorithm that takes at most $\sqrt{n}$ colors, then we prove the above conjecture.

There are also several interesting questions that can be asked about the chromatic number of graphs without certain cycles. We suspect that if $G$ does not have cycles of length $a \bmod k$, then $\chi(G) \leq k + f(k)$, where $f(k) = o(k)$ (possibly a constant). See [13] and [12] for more open problems.